\documentclass[a4paper,11pt]{article}
\pdfoutput=1 % if your are submitting a pdflatex (i.e. if you have
\usepackage{jheppub} % for details on the use of the package, please
\usepackage[T1]{fontenc} % if needed
\usepackage{multirow}%表格合并

\newcommand{\f}{\frac}
\newcommand{\lt}{\left}
\newcommand{\n}{\nonumber}
\newcommand{\p}{\partial}
\newcommand{\dd}{{\rm d}}
\newcommand{\rt}{\right}

\newcommand{\arxgr}[1]{\href{http://arxiv.org/abs/#1}{{\ttfamily arXiv:#1[gr-qc]}}}
\newcommand{\arxth}[1]{\href{http://arxiv.org/abs/#1}{{\ttfamily arXiv:#1[hep-th]}}}

\newcommand{\Arxgr}[1]{\href{http://arxiv.org/abs/gr-qc/#1}{{\ttfamily arXiv:#1[gr-qc]}}}
\newcommand{\Arxth}[1]{\href{http://arxiv.org/abs/hep-th/#1}{{\ttfamily arXiv:#1[hep-th]}}}

\title{\boldmath On the dual relation in the Hawking--Page phase transition of the black holes in a cavity}
\author[a]{Bing-Yu Su}
\author[a,1]{and Nan Li \note{Corresponding author.}}
\affiliation[a]{Department of Physics, College of Sciences, Northeastern University \\ No. 3-11, Wenhua Road, Shenyang, 110819, China}
\emailAdd{linan@mail.neu.edu.cn}
\abstract{The Hawking--Page phase transitions of the $d$-dimensional Schwarzschild and charged black holes are explored in a cavity. The phase transition temperature $T_{\rm HP}$, the minimum black hole temperature $T_0$, and the Gibbs free energy $G$ are systematically calculated. A dual relation for the Schwarzschild black holes in the anti-de Sitter space, $T_{\rm HP}(d)=T_0(d+1)$, is found to be also approximately valid in the cavity case to a high precision, and this relation can be further generalized to the charged black holes in a suitable form. Our work reveals the universal properties of the black holes in different extended phase spaces and motivates further studies on their thermodynamic behaviors that are sensitive to specific boundary conditions, like the terminal points in the $G$--$T$ curves.}

\begin{document}
\maketitle
\flushbottom

\section{Introduction} \label{sec:intro}

Black hole thermodynamics indicates that a black hole is not merely a mathematical singularity, but should be regarded as a complicated physical system with temperature and entropy \cite{Bekenstein, law}. However, there still remain two obvious discrepancies between it and traditional thermodynamics. The first and apparent one is that there is no usual $p$--$V$ term in the first law of black hole thermodynamics. The second and essential one is that a black hole in asymptotically flat space has negative heat capacity and is thus unstable. Actually, these two issues can be handled together. The basic idea is to impose new appropriate boundary conditions and to simultaneously introduce effective pressure and volume to the black hole system. In this way, two new dimensions are added into the black hole phase space, so such theory is named as black hole thermodynamics in the ``extended phase space'' \cite{KM}. Thus, the $p$--$V$ term reappears, and a black hole possesses a stable branch with positive heat capacity, more consistent with traditional thermodynamics.

The gravitational potential of a black hole in asymptotically flat space vanishes at infinity, so the basic function of an appropriate boundary condition is to increase the gravitational potential at large distances. By this means, the boundary plays a role of a reflecting wall against the Hawking radiation \cite{rad} and thus stabilizes the black hole. In principle, any boundary condition meeting such requirements will work. Amongst them, there are two natural choices, corresponding to the two different extended phase spaces discussed in this work.

The first one is to enclose the black hole in the anti-de Sitter (AdS) space, and a positive effective pressure $p$ is introduced from the negative cosmological constant $\Lambda$ as \cite{Kastor, Dolan}
\begin{align}
p=-\f{\Lambda}{8\pi}=\f{(d-1)(d-2)}{16\pi l^2}, \n
\end{align}
where $d$ is the dimension of space-time, and $l$ is the AdS curvature radius. If $\Lambda$ is running, $p$ can be regarded as a thermodynamic quantity, and then an effective black hole volume can be defined as its conjugate variable as $V=({\p M}/{\p p})_S$, with $M$ being the black hole mass \cite{KM}. Moreover, it is found that the $p$--$V$ term in the first law of black hole thermodynamics is $V\,\dd p$ but not $-p\,\dd V$. Therefore, $M$ should be considered as the enthalpy rather than the thermal energy of the black hole.

The second one is to place the black hole still in asymptotically flat space, but to confine it in a spherical cavity, on the wall of which the black hole metric is fixed. Hence, the gravitational potential on the wall is infinite and can play a similar role of the AdS space. In the cavity case, there is a new characteristic length: the cavity radius $r_B$, so we formally introduce an effective black hole volume $V$ as the Euclidean volume,
\begin{align}
V=\f{\Omega r_B^{d-1}}{d-1}, \label{VVV}
\end{align}
where $\Omega$ is the $(d-2)$-dimensional total solid angle. If $r_B$ is allowed to vary, $V$ becomes a thermodynamic quantity, and an effective pressure can be defined as its conjugate variable as $p=-({\p E}/{\p V})_{S}$, with $E$ being the thermal energy of the black hole \cite{Wang}. In this way, another extended phase space can be constructed. (For more relevant works on black hole thermodynamics in a cavity, see Refs. \cite{York, zhaodao, Parentani:1994wr, Gregory:2001bd, Carlip, Lundgren, Akbar:2004ke, Emparan:2012be, Dolansss, 27, Ponglertsakul1, 28, 29, Ponglertsakul2, Sanchis-Gualhhh, 31, 30, Dias11, Dias12, 35, Wang:2019kxp, 36, Liang:2019dni, Wang:2019urm, Wang:2019cax, 37, Wang:2020osg, Tzikas}.)

It is interesting to see that the orders of introduction of the $p$--$V$ term are exactly opposite in the AdS and cavity cases. Consequently, there should be remarkable similarities and notable dissimilarities between them at the same time. For instance, there exists a minimum temperature $T_0$, above which there are two black hole solutions: a stable large black hole and an unstable small black hole, but below which there is no black hole solution at all. The stable large black hole possesses an equation of state like that of a non-ideal fluid, and thus has rich phase transitions and critical phenomena. These topics have been extensively studied in the AdS case (see Ref. \cite{rev} for a review), but the relevant studies on the cavity case are just starting \cite{Wang, zwb}.

One of the most important issues in black hole thermodynamics is the Hawking--Page (HP) phase transition between the stable large black hole and the thermal gas. It was first studied for the Schwarzschild black hole in Ref. \cite{HP} and then for the charged black hole in Refs. \cite{Chamblin1, Chamblin2, Peca}, especially under the influence of the AdS/CFT duality \cite{Witten, Cald, Birmingham:2002ph, Herzog:2006ra, Cai:2007wz, Nicolini:2011dp, Eune:2013qs, Adams:2014vza, Banados:2016hze, Czinner:2017tjq, Aharony:2019vgs, Mejrhit:2019oyi, Lala:2020lge}. After the introduction of the extended phase space in the AdS space, this topic was investigated again with fruitful results \cite{Spallucci:2013jja, Altamirano:2014tva, 1404.2126, Xu:2015rfa, Maity:2015ida, Hansen:2016ayo, Sahay:2017hlq, Mbarek:2018bau, sanren, HPwo, Astefanesei:2019ehu, Wu:2020tmz, DeBiasio:2020xkv, Li:2020khm, Wang:2020pmb, Belhaj:2020mdr, ydw}. However, the corresponding discussions in the extended phase space in a cavity are far from sufficient \cite{zwb}. Currently, the relevant studies are mainly focused on various gravity theories beyond the Einstein gravity.

The most straightforward generalization of the Einstein gravity is the high-dimensional general relativity.
%However, the extension to the $d$-dimensional space-time is not a direct task, and many familiar results will receive nontrivial modifications. For instance, in 5 dimensions, there exist the black ring solutions with one or two angular momenta \cite{Emparan, rus2}. Thus, it is highly necessary to investigate the HP phase transitions in the general $d$-dimensional extended phase spaces.
In Ref. \cite{Wei:2020kra}, the authors discovered an interesting dual relation of the HP phase transition temperature $T_{\rm HP}(d)$ in $d$ dimensions and the minimum black hole temperature $T_0(d+1)$ in $d+1$ dimensions for the Schwarzschild black holes in the AdS space,
\begin{align}
T_{\rm HP}(d)=T_0(d+1). \label{ddddd}
\end{align}
Some possible interpretation was presented, like the duality between the ground and excited states of the black holes \cite{Wei:2020kra}. However, the underlying explanation is still absent, as $T_{\rm HP}$ and $T_0$ are both for the gravitational system, unlike the usual duality from the holographic argument.

The purpose of this work is to study the dual relation in the HP phase transition in the extended phase space in a cavity, and to compare the results to the AdS counterpart, in order to check its universality. First, we derive all the analytical expressions of the HP temperature $T_{\rm HP}$, the minimum black hole temperature $T_0$, and the Gibbs free energy $G$ in arbitrary dimension, which were usually obtained numerically in previous literature. Second, we study the dimension-dependence of $T_{\rm HP}$ and $T_0$, and show that the dual relation in Eq. (\ref{ddddd}) is still approximately valid in the cavity case to a high precision. For the charged black holes, we reformulate Eq. (\ref{ddddd}) in a more concise and convenient way, suitable for both two extended phase spaces. Third, we also discuss the differences between the two extended phase spaces, with emphasis on the terminal points in the $G$--$T$ curves. Altogether, we wish to present a thorough understanding of the HP phase transition and the dual relation in different extended phase spaces to the most general extent.

\section{Black hole thermodynamics in a cavity} \label{sec:thermo}

We start from the black hole thermodynamics in the extended phase space in a cavity. The action of the $d$-dimensional charged black hole in a cavity consists of two parts \cite{Wang:2019urm},
\begin{align}
I&=I_{\rm bulk}+I_{\rm surface}\n\\
&=\f{1}{16\pi}\int_{\cal M}\dd^d x\,\sqrt{-g}(R-F_{\mu\nu}F^{\mu\nu})-\f{1}{16\pi}\int_{\p{\cal M}}\dd^{d-1} x\,\sqrt{-\gamma}\lt[2(K-K_0)+F_{\mu\nu}A^\mu n^\nu\rt], \n
\end{align}
where $\p{\cal M}$ is the boundary of the spherical cavity with a radius $r_B$, $\gamma$ is the determinant of the induced metric on $\p{\cal M}$, $n^\nu$ is the normal vector, $K$ is the trace of the extrinsic curvature tensor, and $K_0$ is its corresponding value when $\p{\cal M}$ is embedded in a flat space. The metric of the spherically symmetric charged black hole is $\dd s^2=-f(r)\,\dd t^2+{\dd r^2}/{f(r)}+r^2\,\dd\Omega^2$, and
\begin{align}
f(r)=1-\f{16\pi M}{(d-2)\Omega r^{d-3}}+\f{32\pi^2 Q^2}{(d-2)(d-3)\Omega^2r^{2d-6}}, \n
\end{align}
where $M$ and $Q$ are the mass and electric charge of the black hole. The horizon radius $r_+$ is determined as the largest root of $f(r)=0$, in terms of which $f(r)$ can be reexpressed as
\begin{align}
f(r)=\lt(1-\f{r_+^{d-3}}{r^{d-3}}\rt)\lt[1-\f{32\pi^2Q^2}{(d-2)(d-3)\Omega^2r_+^{d-3}r^{d-3}}\rt]. \label{lllll}
\end{align}

The thermodynamic properties of the charged black hole can be studied by the Euclidean action $I_{\rm E}$, which is obtained via the analytic continuation method as $I_{\rm E}=iI$. By this means, the Helmholtz free energy of the black hole is obtained as $F=TI_{\rm E}=F(T,Q,r_B;r_+)$. From Eq. (\ref{lllll}), calculating the minimum of $F$ with respect to $r_+$, we have $f'(r_+)=4\pi T\sqrt{f(r_B)}$. Hence, the temperature of the black hole in a cavity is \cite{zhaodao}
\begin{align}
T=\f{1}{\sqrt{f(r_B)}}\lt[\f{d-3}{4\pi r_+}-\f{8\pi Q^2}{(d-2)\Omega ^2r_+^{2d-5}}\rt]. \label{TQ}
\end{align}
Therefore, the temperature $T$ measured at the cavity radius $r_B$ is blue-shifted from the Hawking temperature $T_{\rm H}=f'(r_+)/(4\pi)$ measured at infinity by a factor $1/\sqrt{f(r_B)}$. Moreover, the black hole entropy can be obtained as one quarter of the $(d-2)$-dimensional horizon volume \cite{Bekenstein, law},
\begin{align}
S=\f{\Omega r_+^{d-2}}{4}. \label{Scav}
\end{align}
Thus, the thermal energy $E$ of the black hole is obtained as
\begin{align}
E=F+TS=\f{(d-2)\Omega r_B^{d-3}}{8\pi}\lt[1-\sqrt{f(r_B)}\rt]. \label{Ecav}
\end{align}

In Eq. (\ref{VVV}), we formally define an effective volume $V$ of the spherically symmetric black hole (no matter neutral or charged) in a cavity as the volume of the $(d-1)$-dimensional ball with a radius $r_B$ in the Euclidean space. Obviously, the cavity radius $r_B$ should always be larger than the horizon radius $r_+$,
\begin{align}
r_B>r_+. \n
\end{align}
This constraint will have significant effects on the HP phase transitions in a cavity, especially for the charged black holes, to be explained in Sect. \ref{sec:RN}. Then, the effective pressure $p$ in a cavity is defined as the conjugate variable of $V$,
\begin{align}
p=-\lt(\f{\p E}{\p V}\rt)_{S,Q}=\f{(d-2)(d-3)}{8\pi r_B^2}\lt[\f{2-\f{r_+^{d-3}}{r_B^{d-3}}-\f{32\pi^2Q^2}{(d-2)(d-3)\Omega ^2r_+^{d-3}r_B^{d-3}}}{2\sqrt{f(r_B)}}-1\rt]. \label{pQ}
\end{align}
Next, the electric potential at horizon reads
\begin{align}
\Phi=\lt(\f{\p E}{\p Q}\rt)_{S,V}=\f{4\pi Q\lt(1-\f{r_+^{d-3}}{r_B^{d-3}}\rt)}{(d-3)\Omega r_+^{d-3}\sqrt{f(r_B)}}.\label{PhiQ}
\end{align}

From Eqs. (\ref{VVV}) and (\ref{TQ})--(\ref{PhiQ}), the first law of black hole thermodynamics in the extended phase space in a cavity can be derived as
\begin{align}
\dd E=T\,\dd S-p\,\dd V+\Phi\,\dd Q. \n
\end{align}
Here, we stress that the variable in the first law is the thermal energy $E$ in a cavity, instead of the black hole mass $M$ (i.e., the enthalpy in the AdS space \cite{KM}), but in the limit of $r_B\to\infty$, they naturally coincide.

Now, we turn to the HP phase transition. Due to the increasing gravitational potential at large distances in the extended phase spaces, there exists a minimum temperature $T_0$ for both neutral and charged black holes. Below $T_0$, there is no black hole solution; above $T_0$, there are two black hole branches: a stable large black hole with larger horizon radius and positive heat capacity and an unstable small one on the contrary. Via the Hawking mechanism, the large black hole can exchange matter and energy with the thermal gas and establish thermal equilibrium with the environment. Due to the conservation of charge, the HP phase transition in the black hole--thermal gas system should be studied in a grand canonical ensemble, as the thermal gas is electrically neutral. In other words, the electric potential at horizon is fixed, but the electric charge is allowed to vary. Therefore, the thermodynamic potential of interest in the HP phase transition is the Gibbs free energy $G$.

It will be shown that Gibbs free energy of the large black hole always decreases with temperature, it becomes zero at the HP temperature $T_{\rm HP}$, where the HP phase transition occurs between the large black hole and the thermal gas that always has vanishing Gibbs free energy due to the non-conservation of particle number. Because the phase transition point corresponds to the global minimum of $G$, the criterion of the HP phase transition in the black hole--thermal gas system should be
\begin{align}
G(T_{\rm HP})=0, \label{panju}
\end{align}
and $T_{\rm HP}$ can thus be determined. Below $T_{\rm HP}$, the thermal gas phase with vanishing $G$ is preferred; above $T_{\rm HP}$, the large black hole phase with negative $G$ is preferred, and the thermal gas will collapse into the black hole. In Sects. \ref{sec:Sch} and \ref{sec:RN}, all the expressions of $T_{\rm HP}$ and $T_0$ for the neutral and charged black holes will be analytically calculated in $d$ dimensions, with special attention paid to their ratio and dual relation.

\section{Dual relation of the Schwarzschild black holes} \label{sec:Sch}

In this section, we discuss the HP phase transition and the dual relation of the Schwarzschild black holes in a cavity, and compare the results with the AdS counterpart.

First, in the extended phase space in the AdS space, the HP temperature and the minimum black hole temperature read \cite{Wei:2020kra}
\begin{align}
T_{\rm HP}=2\sqrt{\f{d-2}{d-1}}\sqrt{\f{p}{\pi}}, \quad T_0=2\sqrt{\f{d-3}{d-2}}\sqrt{\f{p}{\pi}}. \label{THP0AdS}
\end{align}
Comparing these results, we have
\begin{align}
\f{T_{\rm HP}(d)}{T_0(d)}=\f{d-2}{\sqrt{(d-1)(d-3)}}. \label{mfffff}
\end{align}
Hence, it is easy to learn $\lim\limits_{d\to\infty}{T_{\rm HP}(d)}/{T_0(d)}=1$, meaning that there will be no metastable large black hole phase when $d\to\infty$. Moreover, in Ref. \cite{Wei:2020kra}, the authors discovered an interesting pressure-independent dual relation of $T_{\rm HP}(d)$ and $T_0(d+1)$,
\begin{align}
T_{\rm HP}(d)=T_0(d+1). \label{SAdS2}
\end{align}
Below, the relations in Eqs. (\ref{mfffff}) and (\ref{SAdS2}) will be further explored for the Schwarzschild black holes in a cavity, in order to check their universality.

In the cavity case, from Eqs. (\ref{VVV}), (\ref{TQ}), (\ref{Ecav}), and (\ref{pQ}), we have the thermal energy, temperature, and pressure as
\begin{align}
E&=\f{(d-2)\Omega r_B^{d-3}}{8\pi}\lt(1-\sqrt{1-\f{r_+^{d-3}}{r_B^{d-3}}}\rt), \label{ESchcav}\\
T&=\f{d-3}{4\pi r_+\sqrt{1-\f{r_+^{d-3}}{r_B^{d-3}}}}, \label{TSchcav}\\
p&=\f{(d-2)(d-3)}{8\pi r_B^2}\lt(\f{2-\f{r_+^{d-3}}{r_B^{d-3}}}{2\sqrt{1-\f{r_+^{d-3}}{r_B^{d-3}}}}-1\rt). \label{pSchcav}
\end{align}
As a result, from Eqs. (\ref{VVV}), (\ref{Scav}), and (\ref{ESchcav})--(\ref{pSchcav}), the Gibbs free energy of the Schwarzschild black hole in a cavity reads
\begin{align}
G=E-TS+pV=\f{(d-2)\Omega r_B^{d-3}}{4\pi(d-1)}\lt[1-\f{4(d-2)-(3d-5) \f{r_+^{d-3}}{r_B^{d-3}}}{4(d-2)\sqrt{1-\f{r_+^{d-3}}{r_B^{d-3}}}}\rt].\label{GSchcav}
\end{align}

When the HP phase transition occurs, from Eqs. (\ref{panju}) and (\ref{GSchcav}), we can first solve $r_B$ in terms of $r_+$. Then, substituting $r_B$ into Eqs. (\ref{TSchcav}) and (\ref{pSchcav}), we obtain the HP temperature $T_{\rm HP}$ as a function of pressure $p$ (i.e., the equation of coexistence line),
\begin{align}
T_{\rm HP}=\lt[\f{(3d-5)^{3d-5}}{4^d(d-1)^{2d-4}(d-2)^{d-1}}\rt]^{\f{1}{2(d-3)}}\sqrt{\f{p}{\pi}}. \label{THPSchcav}
\end{align}
Similar to the AdS case, $T_{\rm HP}$ increases with $p$ and $d$. Since there is no critical point in the coexistence line, the HP phase transition can occur at all pressures, more like a solid--liquid phase transition rather than a liquid--gas one \cite{1404.2126}.

For the minimum temperature $T_0$ of the Schwarzschild black hole in a cavity, the calculation is more difficult than that in the AdS case, because $T$ and $p$ are expressed in the parametric equations in Eqs. (\ref{TSchcav}) and (\ref{pSchcav}). For this reason, from Eq. (\ref{TSchcav}), we first solve $r_B$ as $r_B=r_+\{{u^2}/{[u^2-(d-3)^2]}\}^{\f{1}{d-3}}$, with $u=4\pi r_+ T$. Substituting $r_B$ into Eq. (\ref{pSchcav}), we have
\begin{align}
\f{p}{(d-2)\pi T^2}=\lt\{\f{[u+(d-3)]^2[u-(d-3)]^{2d-4}}{u^{3d-5}}\rt\}^{\f{1}{d-3}}. \n
\end{align}
Therefore, when $u=d-3+2\sqrt{(d-2)(d-3)}$, the minimum temperature $T_0$ can be obtained, which also monotonically increases with $p$ and $d$,
\begin{align}
T_0=\lt\{\f{[d-3+2\sqrt{(d-2)(d-3)}]^{3d-5}}{4^{d-1}(d-2)^{2d-5}(d-3)^{d-1}[2d-5+2\sqrt{(d-2)(d-3)}]}\rt\}^{\f{1}{2(d-3)}}\sqrt{\f{p}{\pi}}. \label{T0Schcav}
\end{align}

To compare with the results in the AdS case, we first calculate the ratio of $T_{\rm HP}$ to $T_0$ in a cavity in $d$ dimensions,
\begin{align}
\f{T_{\rm HP}(d)}{T_0(d)} =\lt\{\f{(d-2)^{d-4}(d-3)^{d-1}(3d-5)^{3d-5}[2d-5+2\sqrt{(d-2)(d-3)}]}{4(d-1)^{2d-4}[d-3+2\sqrt{(d-2)(d-3)}]^{3d-5}}\rt\}^{\f{1}{2(d-3)}}. \label{Tdcav}
\end{align}
Apparently, Eq. (\ref{Tdcav}) is much more complicated than Eq. (\ref{mfffff}). However, by a direct power counting, we easily find in the numerator and denominator that the powers of $d$ are both $5d-9$, and the numerical factors are both $4\cdot 3^{3d-5}$. Hence, we still have the same limit as $\lim\limits_{d\to\infty}{T_{\rm HP}(d)}/{T_0(d)}=1$.

Second, it seems not straightforward to reproduce the simple dual relation $T_{\rm HP}(d)$ $=T_0(d+1)$ in Eq. (\ref{SAdS2}). Nevertheless, we should state that this interesting relation does hold approximately in the cavity case, only with slight deviation. From Eqs. (\ref{THPSchcav}) and (\ref{T0Schcav}), we have
\begin{align}
\f{T_{\rm HP}(d)}{T_0(d+1)}=\lt\{\f{[2d-3+2\sqrt{(d-1)(d-2)}]^{d-3}(3d-5)^{3d^2-11d+10}}{4^d(d-1)^{d-1}(d-2)^2 [d-2+2\sqrt{(d-1)(d-2)}]^{3d^2-11d+6}}\rt\}^{\f{1}{2(d-2)(d-3)}}. \label{TTScav}
\end{align}
Despite the apparent complicity, Eq. (\ref{TTScav}) is still independent of pressure $p$. Again, by a power counting, we find that the ratios of $T_{\rm HP}(d)/T_0(d+1)$ are very close to 1, even when $d$ is only 4. The specific values of $T_{\rm HP}(d)/T_0(d+1)$ for the Schwarzschild black holes in the AdS and cavity cases are both listed in Tab. \ref{t}. We clearly observe that the ratios in the cavity case are precisely close to 1 and approach to 1 when $d\to\infty$. Therefore, we are allowed to conclude that the dual relation in Eq. (\ref{SAdS2}) is truly a remarkable and universal character of the HP phase transitions in different extended phase spaces, insensitive to the boundary conditions of the black holes.
\begin{table}[h]
\centering
\begin{tabular}{c|c|c|c|c|c}
\hline\hline
       & \multicolumn{2}{c|}{AdS} &\multicolumn{3}{c}{cavity}\\
\hline
$d$    & $T_{\rm HP}$ & $T_0$ & $T_{\rm HP}$ & $T_0$ & $\dfrac{T_{\rm HP}(d)}{T_0(d+1)}$ \\
\hline
4      & 0.92132       & 0.79789     & 1.25707       & 1.13393  & 0.99853        \\
\hline
5      & 0.97721       & 0.92132     & 1.31413       & 1.25892  & 0.99930        \\
\hline
6      & 1.00925       & 0.97721     & 1.34658       & 1.31505  & 0.99959        \\
\hline
7      & 1.03006       & 1.00925     & 1.36756       & 1.34713  & 0.99974        \\
\hline
8      & 1.04468       & 1.03006     & 1.38225       & 1.36792  & 0.99982        \\
\hline
9      & 1.05550       & 1.04468     & 1.39311       & 1.38250  & 0.99986        \\
\hline
10     & 1.06385       & 1.05550     & 1.40147       & 1.39330  & 0.99989        \\
\hline
$\infty$ & 1.12838     & 1.12838     & 1.46581       & 1.46581  & 1.00000        \\
\hline\hline
\end{tabular}
\caption{The specific values of the HP temperature $T_{\rm HP}$ and the minimum temperature $T_0$ (normalized by $\sqrt{p}$) of the Schwarzschild black holes in the AdS space and in a cavity, with different dimensions $d$ [see Eqs. (\ref{THP0AdS}), (\ref{THPSchcav}), and (\ref{T0Schcav}) for their expressions]. The ratios of $T_{\rm HP}(d)/T_0(d+1)$ in the cavity case are precisely close to 1 even when $d$ is only 4, and the deviation from 1 vanishes as $d\to\infty$, confirming the universality of the dual relation $T_{\rm HP}(d)=T_0(d+1)$ .} \label{t}
\end{table}

\begin{figure}[h]
\centering
\includegraphics[width=0.6\textwidth]{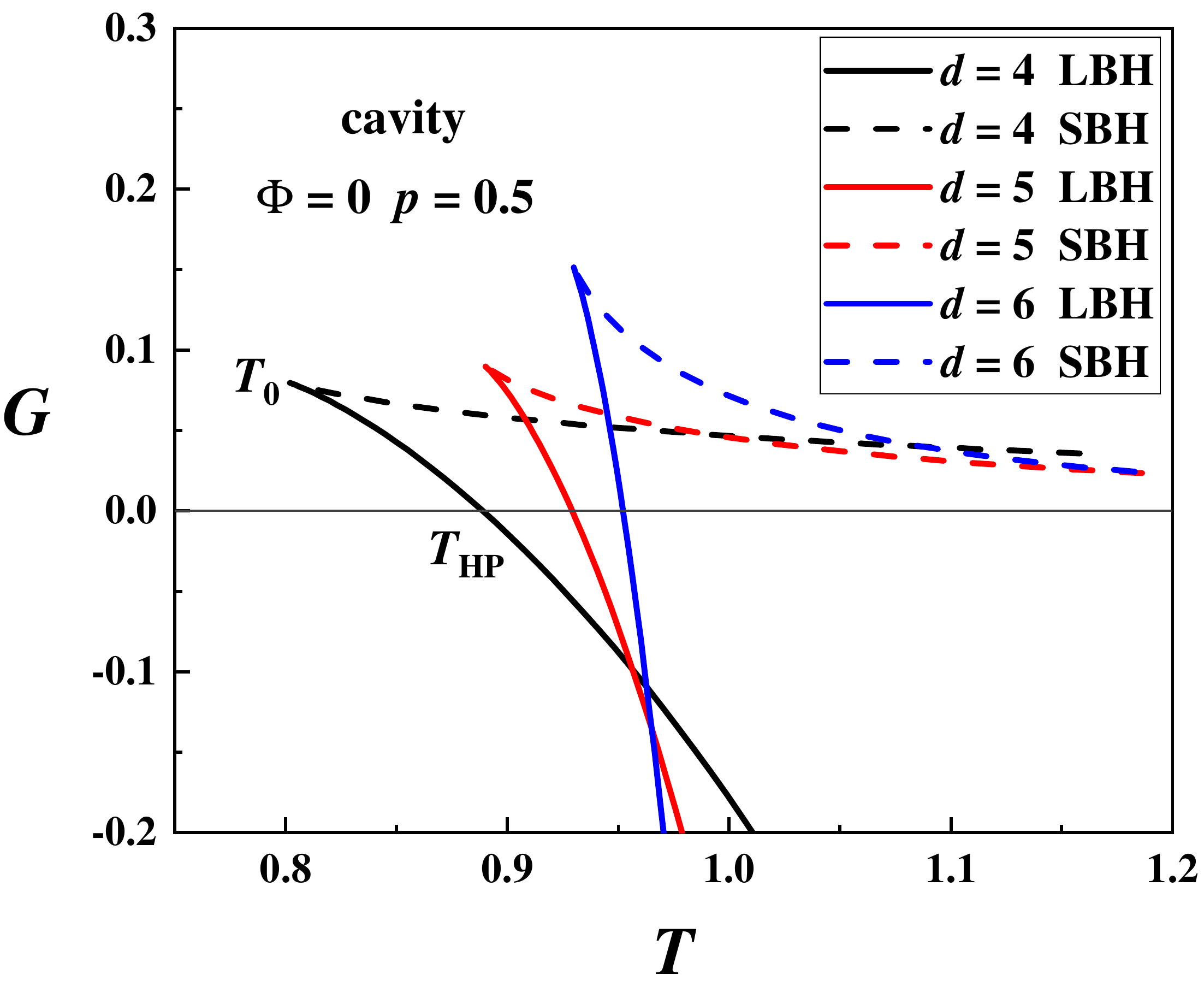}
\caption{The Gibbs free energies $G$ of the Schwarzschild black holes in a cavity as a function of temperature $T$, with pressure $p=0.2$ and different dimensions $d$ (LBH and SBH stand for large and small black hole). There is no HP phase transition for the small black holes. The $G$--$T$ curves of the large black hole phase and the thermal gas phase intersect at the HP temperature $T_{\rm HP}$, corresponding to the first-order HP phase transition. When $d$ increases, the slopes of the $G$--$T$ curves become larger, so the metastable large black hole phase tends to disappear. Analogous to the AdS case, the HP temperature $T_{\rm HP}$ in $d$ dimensions are approximately equal to the minimum black hole temperature $T_0$ in $d+1$ dimensions, as indicated in the dual relation $T_{\rm HP}(d)=T_0(d+1)$.} \label{fig:GT0cavity}
\end{figure}

Finally, the $G$--$T$ curves of the Schwarzschild black holes in a cavity in different dimensions are shown in Fig. \ref{fig:GT0cavity}. The two branches of the curves correspond to the large and small black holes, meeting at $(T_0,G(T_0))$ with a cusp. The $G$--$T$ curves of the small black holes are concave and will never reach the $T$-axis, so there is no HP phase transition for them. On the contrary, the $G$--$T$ curves of the large black holes intersect the $T$-axis at $T_{\rm HP}$. As the derivatives of the $G$--$T$ curves are discontinuous at $T_{\rm HP}$, the HP phase transition is of first-order. Furthermore, we clearly observe that the slopes of the $G$--$T$ curves of the large black holes become larger in high dimensions. This means that the metastable large black hole phase [i.e., the segment of the $G$--$T$ curve in the temperature interval $(T_0,T_{\rm HP})$] tends to disappear when $d$ increases, confirming the result in Eq. (\ref{Tdcav}). Moreover, the HP temperature $T_{\rm HP}$ in $d$ dimensions are approximately equal to the minimum temperature $T_0$ in $d+1$ dimensions, indicating that the dual relation $T_{\rm HP}(d)=T_0(d+1)$ is a universal property in different extended phase spaces.

\section{Dual relation of the charged black holes} \label{sec:RN}

In this section, we discuss the HP phase transition and the dual relation of the charged black holes in a cavity, and the influences from the electric potential will be studied carefully.

Again, we first summarize the relevant results in the AdS space for reference. The HP temperature $T_{\rm HP}$ and the minimum black hole temperature $T_0$ read
\begin{align}
T_{\rm HP}&=2\sqrt{\f{d-2-2(d-3)\Phi^2}{d-1}}\sqrt{\f{p}{\pi}}, \label{THPRNAdS}\\
T_0&=\f{2\sqrt{(d-3)[d-2-2(d-3)\Phi^2]}}{d-2}\sqrt{\f{p}{\pi}}. \label{T0RNAdS}
\end{align}
It is straightforward to find that both $T_{\rm HP}$ and $T_0$ monotonically decrease with $\Phi$. Moreover, if the HP phase transition occurs, as $T_{\rm HP}$ is positive, from Eq. (\ref{THPRNAdS}), the electric potential $\Phi$ must have an upper bound,
\begin{align}
\Phi<\sqrt{\f{d-2}{2(d-3)}}. \label{yueshu}
\end{align}

For the relation of $T_{\rm HP}$ and $T_0$, from Eqs. (\ref{THPRNAdS}) and (\ref{T0RNAdS}), we have ${T_{\rm HP}(d)}/{T_0(d)} =(d-2)/{\sqrt{(d-1)(d-3)}}$. Notably, this ratio is the same as that for the Schwarzschild black hole in Eq. (\ref{mfffff}), as the dependence on $\Phi$ cancels out. Hence, $\Phi$ does not affect the ratio of ${T_{\rm HP}(d)}/{T_0(d)}$, although it does decrease $T_{\rm HP}$ and $T_0$ individually. Hence, we still have $\lim\limits_{d\to\infty}{T_{\rm HP}(d)}/{T_0(d)}=1$. Furthermore, the simple dual relation $T_{\rm HP}(d)=T_0(d+1)$ for the Schwarzschild black hole in Eq. (\ref{SAdS2}) no longer holds in the current circumstance. At present, the ratio of $T_{\rm HP}(d)/T_0(d+1)$ depends on both $d$ and $\Phi$,
\begin{align}
\f{T_{\rm HP}(d)}{T_0(d+1)}=\sqrt{\f{(d-1)[d-2-2(d-3)\Phi^2]}{(d-2)[d-1-2(d-2)\Phi^2]}}. \n
\end{align}
In Ref. \cite{Wei:2020kra}, the authors generalized the dual relation in Eq. (\ref{SAdS2}) to the charged black hole in the AdS space as
\begin{align}
T_{\rm HP}(d,\Phi)=T_0\lt(d+1,\f{\sqrt{(d-1)(d-3)}}{d-2}\Phi\rt). \label{fmmmm}
\end{align}
Whereas, we point out that, from the ratio of ${T_{\rm HP}(d)}/{T_0(d)}$, the dual relation in Eq. (\ref{fmmmm}) can actually be rewritten in a more concise form,
\begin{align}
T_{\rm HP}(d,\Phi)=T_0\lt(d+1,\f{T_0(d)}{T_{\rm HP}(d)}\Phi\rt). \label{RNAdSdual}
\end{align}
We must emphasize that the essential reason why we reformulate Eq. (\ref{fmmmm}) as Eq. (\ref{RNAdSdual}) is not merely for mathematical tidiness, but for deeper physical consideration. It will be shown that Eq. (\ref{RNAdSdual}) still holds approximately for the charged black hole in a cavity, albeit the ratio of ${T_0(d)}/{T_{\rm HP}(d)}$ becomes much more complicated. In other words, it is the dual relation expressed in the form of Eq. (\ref{RNAdSdual}) that can be applied to the cavity case, not the original one in Eq. (\ref{fmmmm}).

Now, we discuss the HP phase transition and the dual relation of the charged black holes in a cavity. This is not just a trivial extension of the results for the Schwarzschild black holes. There exist significant differences, especially in the $G$--$T$ curves. Following the same procedure in Sect. \ref{sec:Sch}, the Gibbs free energy of the charged black hole in a cavity reads
\begin{align}
G&=E+pV-TS-\Phi Q=\f{(d-2)\Omega r_B^{d-3}}{4\pi(d-1)} \n\\
&\quad\times \lt\{1-\sqrt{\f{d-2}{d-2-(d-2-2(d-3)\Phi^2)\f{r_+^{d-3}}{r_B^{d-3}}}}\lt[1-\f{(3d-5)(d-2-2(d-3)\Phi^2) \f{r_+^{d-3}}{r_B^{d-3}}}{4(d-2)^2}\rt]\rt\}. \label{GRNcav}
\end{align}
If the HP phase transition occurs, setting $G=0$ in Eq. (\ref{GRNcav}), we can solve the cavity radius $r_B$ in terms of the horizon radius $r_+$,
\begin{align}
r_B=r_+\lt\{\f{(3d-5)^2[d-2-2(d-3)\Phi^2]}{8(d-1)(d-2)^2}\rt\}^{\f{1}{d-3}}, \n
\end{align}
and the constraint $r_B>r_+$ immediately sets an upper bound of $\Phi$ in a cavity,
\begin{align}
\Phi<\sqrt{\f{(d-2)(d-3)}{2(3d-5)^2}}. \label{PhiRNcav}
\end{align}
Comparing the results in Eqs. (\ref{yueshu}) and (\ref{PhiRNcav}), we find that this upper bound is much stricter than that in the AdS case.

According to the method in Sect. \ref{sec:Sch}, after some algebra, the HP temperature can be obtained as
\begin{align}
T_{\rm HP}=\lt\{\f{(3d-5)^{3d-5}[d-2-2(d-3)\Phi^2]^{2d-4}}{4^d(d-1)^{2d-4}(d-2)^{3d-5}}\rt\}^{\f{1}{2(d-3)}}\sqrt{\f{p}{\pi}}.\n
\end{align}
Here, we mention that the requirement $T_{\rm HP}>0$ sets the same loose constraint on $\Phi$ as that in Eq. (\ref{yueshu}), so the tighter upper bound in Eq. (\ref{PhiRNcav}) will not change. Furthermore, the minimum temperature $T_0$ of the charged black hole reads
\begin{align}
T_0=\lt\{\f{[d-3+2\sqrt{(d-2)(d-3)}]^{3d-5}[d-2-2(d-3)\Phi^2]^{2d-4}}{4^{d-1}(d-2)^{4d-9}(d-3)^{d-1} [2d-5+2\sqrt{(d-2)(d-3)}]}\rt\}^{\f{1}{2(d-3)}}\sqrt{\f{p}{\pi}}. \n
\end{align}
Similar to the AdS case, both $T_{\rm HP}$ and $T_0$ decrease with $\Phi$. However, different from the AdS case, they both increase with $d$ for any given value of $\Phi$, as the upper bound of $\Phi$ is so small.

Now, we discuss the relation of $T_{\rm HP}$ and $T_0$ for the charged black hole in a cavity. First, despite the formal complicities of $T_{\rm HP}(d)$ and $T_0(d)$, it is interesting to find that their ratio is exactly the same as that in Eq. (\ref{Tdcav}) for the Schwarzschild black hole in a cavity. Consequently, we are allowed to conclude that the independence of ${T_{\rm HP}(d)}/{T_0(d)}$ on the electric potential $\Phi$ is universal, valid in both the AdS and cavity cases. Therefore, as in Eq. (\ref{Tdcav}), we still have $\lim\limits_{d\to\infty}{T_{\rm HP}(d)}/{T_0(d)}=1$.

\begin{figure}[h]
\centering % \begin{center}/\end{center} takes some additional vertical space
\includegraphics[width=0.6\textwidth]{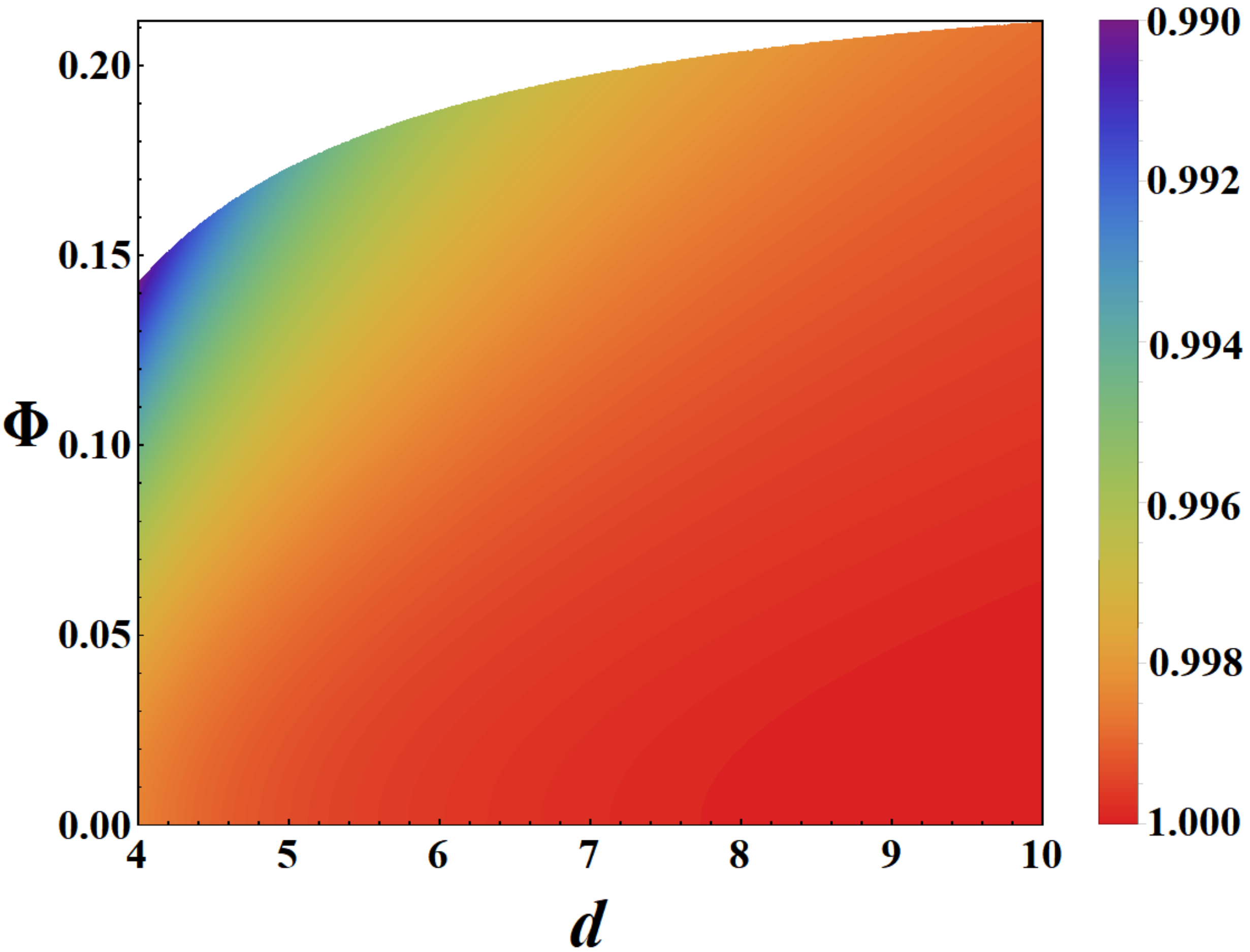}
\caption{The ratio of $T_{\rm HP}(d,\Phi)/T_0(d+1,\f{T_0(d)}{T_{\rm HP}(d)}\Phi)$ of the charged black hole in a cavity as a function of dimension $d$ and electric potential $\Phi$. The upper bound of $\Phi$ is also plotted according to Eq. (\ref{PhiRNcav}). The dual relation rewritten in the form of Eq. (\ref{RNAdSdual}) is still precisely valid in the cavity case. The ratio is already $0.99092$ when $d=4$ and $\Phi=1/7$, and will be even closer to 1 with larger $d$ and smaller $\Phi$.} \label{fig:dFTT}
\end{figure}

Second, for the dual relation of $T_{\rm HP}(d)$ and $T_0(d+1)$, we first point out that the original one in Eq. (\ref{fmmmm}) suggested in Ref. \cite{Wei:2020kra} is no longer suitable for the current discussion, as the ratio of $\sqrt{(d-1)(d-3)}/(d-2)$ in the AdS case has no direct physical interpretation in the cavity case. However, we will show that the dual relation rewritten in the form of Eq. (\ref{RNAdSdual}) does apply to the charged black holes in a cavity rather well. In Fig. \ref{fig:dFTT}, we illustrate the influences of $d$ and $\Phi$ on the ratio of
\begin{align}
\f{T_{\rm HP}(d,\Phi)}{T_0\lt(d+1,\f{T_0(d)}{T_{\rm HP}(d)}\Phi\rt)}. \label{bds}
\end{align}
The explicit expression of Eq. (\ref{bds}) is not shown due to the unnecessary mathematical tediousness. We clearly see that this ratio is always approximately equal to 1, and will be even closer to 1 with larger $d$ and smaller $\Phi$. These observations further support that our new dual relation in Eq. (\ref{RNAdSdual}) is suitable for both the AdS and cavity cases, confirming its universality in different extended phase spaces.

Last, we mention a distinctive character of the $G$--$T$ curves of the charged black holes in a cavity. Due to the extra constraint $r_B>r_+$ in a cavity, there will be terminal points in the $G$--$T$ curves, which are absent in the AdS case. To figure out these terminal points, from Eq. (\ref{pQ}), we first solve $r_+$ in terms of $r_B$ as
\begin{align}
r_+=r_B\lt\{\f{8\sqrt{\pi pr_B^2[4\pi pr_B^2+(d-2)(d-3)]}}{(d-2)(d-3)^2[d-2-2(d-3)\Phi^2]}\lt[\sqrt{4\pi pr_B^2+(d-2)(d-3)}-\sqrt{4\pi pr_B^2}\rt]^2\rt\}^{\f{1}{d-3}}. \label{r+rBp}
\end{align}
Then, considering the constraint $r_B>r_+$, we obtain the upper bound of $r_B$ as
\begin{align}
r_B<\f{[2(d-2)(d-3)]^{\f14}[\sqrt{d-2}-\sqrt{2(d-3)}\Phi]}{4\sqrt{2\pi p\Phi}}. \label{up}
\end{align}
This is evidently different from the case of the Schwarzschild black hole in a cavity, in which $r_B$ can take any value. Nevertheless, in the limit of $\Phi\to 0$, the upper bound of $r_B$ naturally tends to infinity. Substituting Eq. (\ref{r+rBp}) into Eqs. (\ref{TQ}) and (\ref{GRNcav}), we are able to reexpress the temperature and Gibbs free energy in the parametric equations as $T=T(d,p,\Phi,r_B)$ and $G=G(d,p,\Phi,r_B)$. Taking Eq. (\ref{up}) into account, we eventually determine the terminal points in the $G$--$T$ curves of the charged black holes in a cavity. Here, we only give their coordinates in 4 dimensions (the general results in $d$ dimensions will not be shown for their lengthy expressions),
\begin{align}
(T,G)=\lt((1+\Phi)\sqrt{\f{p}{2\pi\Phi}},-\f{(1-\Phi^2)(1-7\Phi)}{24\sqrt{2\pi p\Phi^3}}\rt).\n
\end{align}

\begin{figure}[h]
\centering % \begin{center}/\end{center} takes some additional vertical space
\includegraphics[width=.45\textwidth]{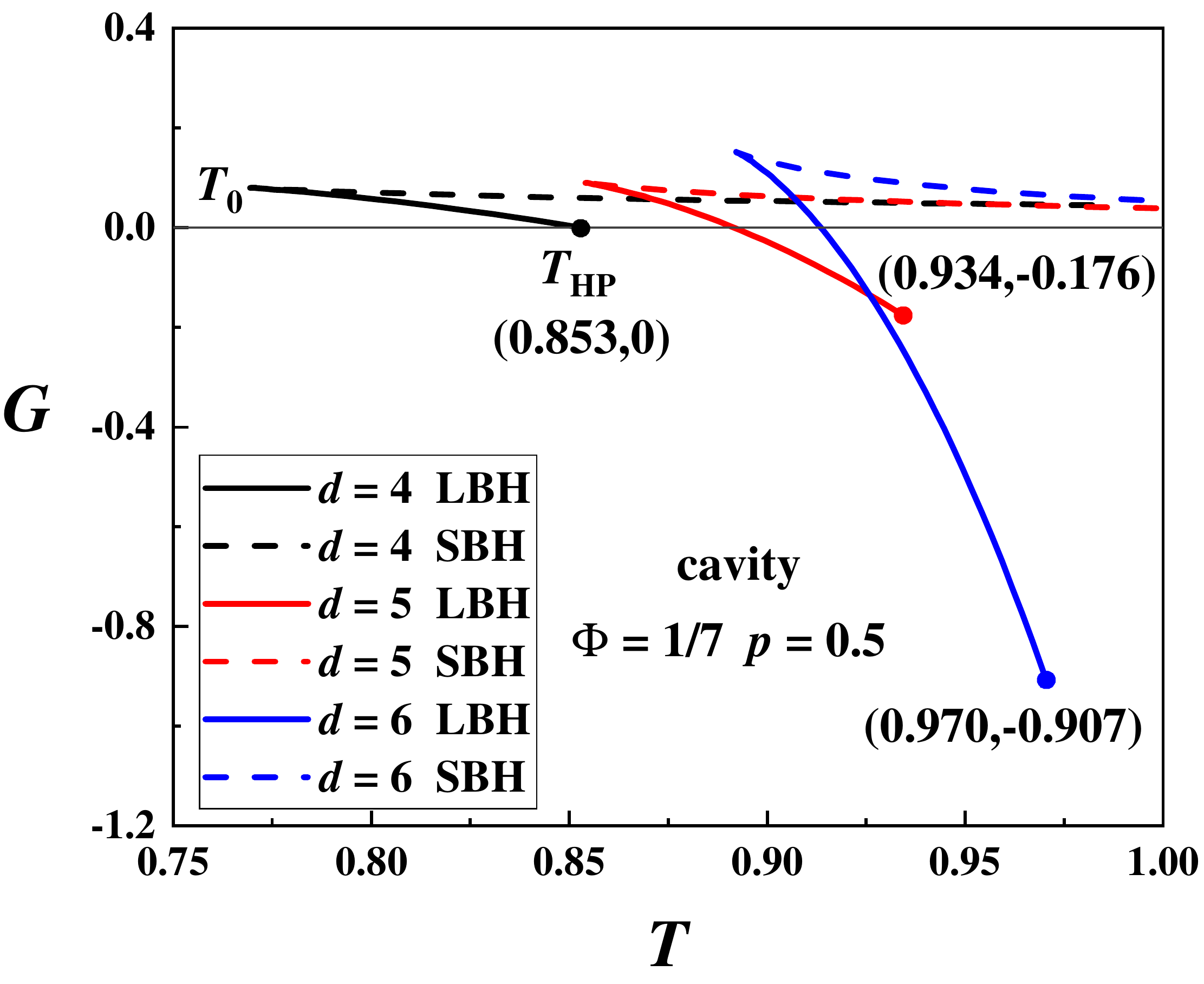} \quad
\includegraphics[width=.45\textwidth]{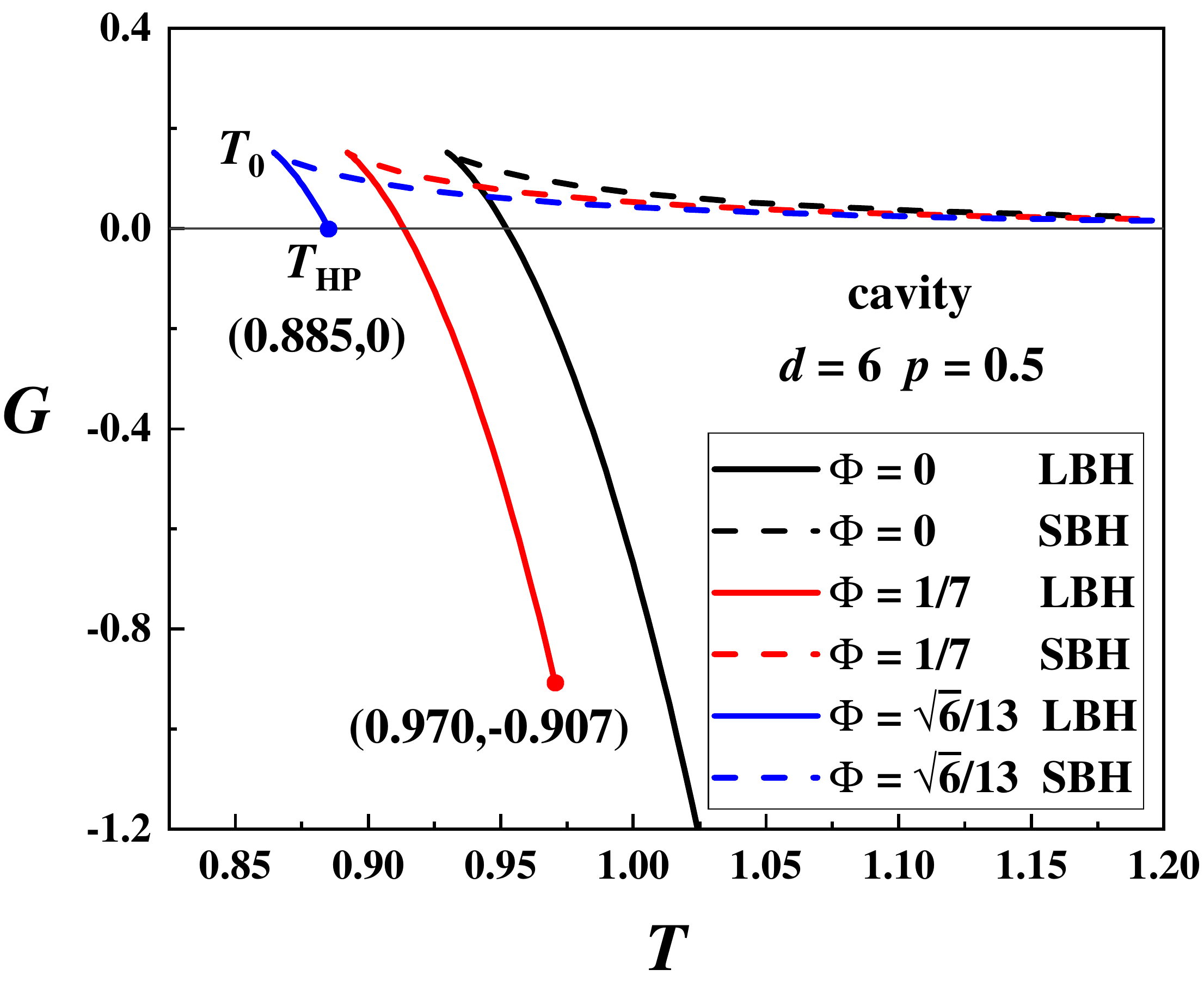}
\caption{The Gibbs free energies $G$ of the charged black holes in a cavity as a function of temperature $T$, with pressure $p=0.5$ and different dimensions $d$ (left panel) and different electric potentials $\Phi$ (right panel). Both $T_{\rm HP}$ and $T_0$ increase with $d$ and decrease with $\Phi$. There are terminal points in the $G$--$T$ curves when $\Phi\neq 0$, and when $\Phi$ reaches its upper bound in Eq. (\ref{PhiRNcav}), the terminal point is located exactly on the $T$-axis at $(T_{\rm HP}, 0)$. Besides, in the lower panel, all the cusps of the $G$--$T$ curves have the same height.} \label{fig:GTRNcavity}
\end{figure}

The $G$--$T$ curves of the charged black holes in a cavity are plotted in Fig. \ref{fig:GTRNcavity}, with different dimensions $d$ and different electric potentials $\Phi$. Because of the terminal point, the temperature of the large charged black hole has an upper bound. In particular, in the right panel, as $\Phi$ increases, the terminal points move toward the upper left corner of the $G$--$T$ plane, and finally when $\Phi$ reaches its upper bound in Eq. (\ref{PhiRNcav}), the terminal point is located exactly on the $T$-axis at $(T_{\rm HP}, 0)$. Moreover, when $\Phi\to 0$, the terminal points move toward the lower right corner of the $G$--$T$ plane, and finally disappear when $\Phi=0$, consistent with the case of the Schwarzschild black hole in a cavity in Fig. \ref{fig:GT0cavity}. Another notable character in the right panel is that all the cusps of the $G$--$T$ curves have the same height. This is because when we substitute Eq. (\ref{r+rBp}) into Eq. (\ref{GRNcav}), the electric potential $\Phi$ cancels out. As a result, when $d$, $p$, and $r_B$ are fixed, except for the different terminal points, all the $G$--$T$ curves distinguish one another only by a shift in the $T$-direction.

\section{Conclusion}\label{sec:con}

In recent years, black hole thermodynamics in the extended phase space is receiving increasing research interests. The basic motivations of these studies are to stabilize the black hole and to restore the $p$--$V$ term in the first law of black hole thermodynamics. There are two complementary choices to realize these aims: in the AdS space and in a cavity. Thus, we achieve two extended phase spaces with different boundary conditions. As a result, there will be similarities and dissimilarities simultaneously. In this work, we systematically investigate the HP phase transition in the two extended phase spaces. The HP temperature $T_{\rm HP}$, the minimum black hole temperature $T_0$, and the Gibbs free energy $G$ are all calculated and illustrated in arbitrary dimensions. In particular, the ratio and dual relation of $T_{\rm HP}$ and $T_0$ are studied in depth. The basic conclusions of our work can be drawn as follows.

1. We find evident similarities in the two extended phase spaces. (1) The HP phase transition can occur at all pressures, and $T_{\rm HP}$ and $T_0$ increase with pressure $p$ in all circumstances. (2) $T_{\rm HP}$ and $T_0$ increase with dimension $d$ for the Schwarzschild black holes, and decrease with electric potential $\Phi$ if the black holes are charged.

2. As for the relations of $T_{\rm HP}$ and $T_0$, the main aspects are similar in the two extended phase spaces. (1) The ratios of $T_{\rm HP}(d)/T_0(d)$ are always independent of $p$ and $\Phi$, and approach to 1 when $d\to\infty$, meaning that the metastable large black hole phase disappears in higher dimensions. (2) For the Schwarzschild black holes, we find that the simple dual relation $T_{\rm HP}(d)=T_0(d+1)$ in the AdS case is also approximately valid in the cavity case to a high precision, and the deviation tends to vanish as $d$ increases. (3) For the charged black holes, we reformulate the dual relation from Eq. (\ref{fmmmm}) suggested in Ref. \cite{Wei:2020kra} to Eq. (\ref{RNAdSdual}). Besides mathematical conciseness, this new form can be further applied to the cavity case in a more reasonable sense.

3. For the charged black holes, there are significant dissimilarities in the two extended phase spaces. (1) Due to the constraint of $r_B>r_+$, if the HP phase transition occurs, the upper bound of $\Phi$ in the cavity case is much tighter than that in the AdS case. (2) There are terminal points in the $G$--$T$ curves in the cavity case. These points are located at $(T_{\rm HP},0)$ when $\Phi$ reaches its upper bound, and will be absent when $\Phi\to 0$.

4. A side remark is that all the results in this work are analytical, which were usually obtained numerically, so that we can have a clear and quantitative understanding of the different extended phase spaces.

In summary, on the one hand, we find remarkable analogies in the HP phase transition and the dual relation in the AdS space and in a cavity. On the other hand, we also observe notable differences at the same time. Generally speaking, all these dissimilarities stem from the fact that the cavity radius $r_B$ must always be larger than the horizon radius $r_+$. Therefore, there are still some subtleties in black hole thermodynamics that are sensitive to the specific boundary conditions. Altogether, we wish to provide a whole picture of the HP phase transition and the dual relation and to motivate further studies on other properties of different extended phase spaces.

\acknowledgments

We are very grateful to Y.-X. Liu, P. Wang, Y.-Y. Wang, Z.-Z. Wang, and S.-W. Wei for fruitful discussions. This work is supported by the Fundamental Research Funds for the Central Universities of China (No. N170504015).

%\paragraph{Note added.} This is also a good position for notes added after the paper has been written.

%\bibliography{mybibfile}

\end{document}